\documentclass[prd,aps,preprint,tightenlines,superscriptaddress]{revtex4}
\usepackage{amssymb}
\usepackage{epsfig}

\preprint{DOE/ER/40762-323} \preprint{UM-PP\#05-008}

\begin{document}

\title{Deuteron Compton Scattering in Effective Field Theory \\
And Spin-Independent Nucleon Polarizabilities}
\author{Jiunn-Wei Chen}
\email{jwc@phys.ntu.edu.tw}
\affiliation{Department of Physics and National Center for Theoretical Sciences at
Taipei, National Taiwan University, Taipei, Taiwan 10617}
\author{Xiangdong Ji}
\email{xji@physics.umd.edu}
\affiliation{Department of Physics, University of Maryland, College Park, Maryland 20742}
\author{Yingchuan Li}
\email{yli@physics.umd.edu}
\affiliation{Department of Physics, University of Maryland, College Park, Maryland 20742}
\date{\today }

\begin{abstract}
Deuteron Compton scattering is calculated to $\mathcal{O}(Q^2)$ in pionless
effective field theory using a dibaryon approach. The vector amplitude,
which was not included in the previous pionless calculations, contributes to
the cross section at $\mathcal{O}(Q^2)$ and influences significantly the
extracted values of nucleon electric polarizability at incident photon
energy 49 MeV. We recommend future high precision deuteron compton
scattering experiments being performed at 25-35 MeV photon energy where the
nucleon polarizability effects are appreciable and the pionless effective
field theory is most reliable. For example, a measurement at 30 MeV with a
3\% error will constrain the isoscalar nucleon electric polarizability $%
\alpha _{0}$ with a $\sim 30\%$ error.
\end{abstract}

\maketitle

\section{Introduction}

Nucleon polarizabilities are fundamental properties of nucleons. They
characterize how easily the nucleons deform under external electromagnetic
fields. For proton polarizabilities, tight constraints have been obtained by
Compton scattering \cite{pdata,mainz,beane} and photoproduction sum rule
\cite{mainz,beane}. For neutron polarizabilities, the extraction is more
complicated and less satisfactory. Scattering neutrons off the Coulomb field
of a heavy nucleus could in principle determine the neutron electric
polarizability \cite{an1,an2,an3,an4}. However, the error is still quite
large. By far the best methods known to measure the neutron polarizabilities
involve photon scattering on deuteron targets. This introduces theoretical
complications such as final-state interactions and two-body currents.

Recently quasi-free Compton scattering ($\gamma d\rightarrow \gamma np$) at
200 to 400 MeV photon energy was carried out at Mainz \cite{kossert02}.
Using the model developed in Ref. \cite{levchuk94}, tight constraints were
reported on the neutron electric and magnetic polarizabilities: $\alpha
_{n}=12.5\pm 1.8(\mathrm{stat}){}_{-0.6}^{+1.1}{}(\mathrm{syst})\pm 1.1(%
\mathrm{model})$ and $\beta _{n}=~\,2.7\mp 1.8(\mathrm{stat}%
){}_{-1.1}^{+0.6}{}(\mathrm{syst})\mp 1.1(\mathrm{model})$ in units of 10$%
^{-4}$ fm$^{3}$ (which will be used throughout this paper) \cite{kossert02}.
The other approach, deuteron Compton scattering ($\gamma d\rightarrow \gamma
d$), was carried out at Illinois \cite{Illinois}, SAL (Saskatoon) \cite{SAL}
and Lund \cite{Lund} with photon energy from 49 to 95 MeV. Several
theoretical calculations are performed using potential models \cite%
{p1,p2,p3,p4}. The latest calculation using the Lund data at 55 and 66 MeV
gives the iso-scalar combinations of nucleon polarizabilities $\alpha
_{0}+\beta _{0}=17.4\pm 3.7$ and $\alpha _{0}-\beta _{0}=6.4\pm 2.4$ which,
when combined with the proton values, imply $\alpha _{n}=~8.8~\pm ~2.4$%
(total) $\pm ~3.0$(model) and $\beta _{n}=~6.5~\mp ~2.4$(total) $\mp ~3.0$%
(model).

An alternative method to analyze deuteron Compton scattering is nuclear
effective field theory (EFT) \cite{Weinberg,KSW}\ (see Ref. \cite{NEFTrev}
for a recent review). The goal of the nuclear EFT program is to establish
model independent, systematic and controlled expansions for few-body and
eventually nuclear matter problems. For systems with characteristic momenta
below the pion mass, theories with pions integrated out are highly
successful and applied to many observables \cite%
{KSW,kaplan,vK97,Cohen97,BHvK1,chen1}. For systems with characteristic
momenta above the pion mass, power counting is more complicated \cite%
{towards}. But for practical purposes, the power counting developed by
Weinberg \cite{Weinberg}\ could still be quite accurate despite its
renormalization problem in the $^{1}S_{0}$ channel. EFT calculations of
deuteron Compton scattering were carried out in Refs. \cite%
{chen2,beane,dEFT,rupak}. The latest extraction of nucleon polarizabilities
using Weinberg's theory but with phenomenological wave functions yields $%
\alpha _{0}=8.9\pm 1.5_{-0.9}^{+4.7}$ and $\beta _{0}=2.2\pm
1.5_{-0.9}^{+1.2}$\cite{beane}. In comparison, using a theory with pion
integrated out, Ref. \cite{rupak} extracts $\alpha _{0}=8.4\pm 3.4$ and $%
\beta _{0}=8.9\pm 4.3$ from the 49 MeV data. Other calculations \cite%
{chen2,dEFT} showed that data below 70 MeV give results consistent with the
values predicted by leading-order chiral perturbation theory $\alpha
_{0}=10\beta _{0}=12$ \cite{meissner}.

In the future, the High-Intensity Gamma Source (HIGS) at Duke University
will be able to measure deuteron Compton scattering with high precision.
Thus it is timely to ask what the best strategy is to improve the
determination of nucleon polarizabilities. In this work, we focus on the low
energy experiments---because the relevant theory is most well
understood---and explore the sensitivity of $\alpha _{0}$ and $\beta _{0}$
in future experiments. We follow the power counting in \cite{dEFT}, and
present the complete unpolarized deuteron Compton scattering cross section
to $\mathcal{O}(Q^2)$ in the pionless dibaryon EFT \cite{kaplan,BHvK1,dEFT}.
A vector amplitude, which was not included in the previous pionless
calculations was found to contribute at the same order as the nucleon
polarizability contribution. The impact of the new contribution to the
extraction of nucleon polarizabilities $\alpha _{0}$ and $\beta _{0}$ is
studied.

\section{Kinematics}

The number of independent structures in Compton scattering amplitudes can be
conveniently analyzed in the helicity basis. In this basis, amplitudes are
characterized by $\left\langle h_{1}^{\prime },h_{2}^{\prime
}|h_{1},h_{2}\right\rangle $, where $h_{i}$($h_{i}^{\prime })$ is the
helicity for particle $i$ in the initial(final) state. Under parity
transformation (P), a helicity amplitude transforms as%
\begin{equation}
P:\left\langle h_{1}^{\prime },h_{2}^{\prime }|h_{1},h_{2}\right\rangle
\rightarrow \left\langle -h_{1}^{\prime },-h_{2}^{\prime
}|-h_{1},-h_{2}\right\rangle \ .
\end{equation}%
While under time reversal transformation (T),%
\begin{equation}
T:\left\langle h_{1}^{\prime },h_{2}^{\prime }|h_{1},h_{2}\right\rangle
\rightarrow \left\langle h_{1},h_{2}|h_{1}^{\prime },h_{2}^{\prime
}\right\rangle \ .
\end{equation}%
It is clear that only the linear combinations of the helicity amplitudes
that are invariant under P and T can contribute to Compton scattering. It is
easy to see that there are $2\left( J+1\right) (2J+1)$ independent helicity
amplitudes for a spin-J target. Thus there are two independent amplitudes
for a spin-0 target, six for a spin-1/2 target, and twelve for a spin-1
target, such as deuteron. The twelve amplitude structures for deuteron
Compton scattering can be further classified as the scalar, vector and
tensor amplitudes, $S$, $V$ and $T$:
\begin{equation}
A=\frac{ie^{2}}{2M_{N}}\left[ S\mathbf{\epsilon }_{d}^{\prime ^{\ast }}\cdot
\mathbf{\epsilon }_{d}+\varepsilon _{ijk}V_{i}\mathbf{\epsilon }%
_{d_{j}}^{\prime ^{\ast }}\mathbf{{}\epsilon }_{d_{k}}+T_{ij}\left( \mathbf{%
\epsilon }_{d_{i}}^{\prime ^{\ast }}\mathbf{\epsilon }_{d_{j}}+\mathbf{%
\epsilon }_{d_{i}}\mathbf{\epsilon }_{d_{j}}^{\prime ^{\ast }}-\frac{2}{3}%
\delta _{ij}\mathbf{\epsilon }_{d}^{\prime ^{\ast }}\cdot \mathbf{\epsilon }%
_{d}\right) \right] ,  \label{A}
\end{equation}%
where $\mathbf{\epsilon }_{d}$ ($\mathbf{\epsilon }_{d}^{\prime }$) is the
initial(final) deuteron polarization, $e$ is the proton charge and $M_{N}$
is the nucleon mass. By the same counting discussed above, $S$, $V$ and $T$
have two, four and six independent structures, respectively. We chose the
following basis for these amplitudes \cite{vecCompton}:
\begin{equation}
S=f_{1}\mathbf{\epsilon }^{\prime }{}^{\ast }\cdot \mathbf{\epsilon }+f_{2}%
\mathbf{s}^{\prime }{}^{\ast }\cdot \mathbf{s\ ,}
\end{equation}%
\begin{equation}
V_{i}=f_{3}(\mathbf{\epsilon }^{\prime }{}^{\ast }\times \mathbf{\epsilon }%
)_{i}+f_{4}(\mathbf{s}^{\prime }{}^{\ast }\times \mathbf{s})_{i}+f_{5}\left(
\mathbf{s}^{\prime }{}^{\ast }\cdot \mathbf{\epsilon \hat{k}}_{i}-\mathbf{%
\epsilon }^{\prime }{}^{\ast }\cdot \mathbf{s\hat{k}}_{i}^{\prime }\right)
+f_{6}\left( \mathbf{s}^{\prime }{}^{\ast }\cdot \mathbf{\epsilon \hat{k}}%
_{i}^{\prime }-\mathbf{\epsilon }^{\prime }{}^{\ast }\cdot \mathbf{s\hat{k}}%
_{i}\right) \ ,
\end{equation}%
\begin{eqnarray}
T_{ij} &=&f_{7}\mathbf{\epsilon }_{_{i}}^{\prime ^{\ast }}\mathbf{\epsilon }%
_{_{j}}+f_{8}\mathbf{s}_{_{i}}^{\prime ^{\ast }}\mathbf{s}%
_{_{j}}+f_{9}\left( \mathbf{\epsilon }^{\prime }{}^{\ast }\cdot \mathbf{\hat{%
k}\hat{k}}_{i}^{\prime }\mathbf{\epsilon }_{_{j}}+\mathbf{\epsilon }\cdot
\mathbf{\hat{k}}^{\prime }\mathbf{\hat{k}}_{i}\mathbf{\epsilon }_{j}^{\prime
}{}^{\ast }\right) +f_{10}\left( \mathbf{s}^{\prime }{}^{\ast }\cdot \mathbf{%
\hat{k}\hat{k}}_{i}^{\prime }\mathbf{s}_{_{j}}+\mathbf{s}\cdot \mathbf{\hat{k%
}}^{\prime }\mathbf{\hat{k}}_{i}\mathbf{s}_{j}^{\prime }{}^{\ast }\right)
\nonumber \\
&&+f_{11}\frac{1}{2}\left( \mathbf{\hat{k}}_{i}\mathbf{\hat{k}}_{j}+\mathbf{%
\hat{k}}_{i}^{\prime }\mathbf{\hat{k}}_{j}^{\prime }\right) \mathbf{\epsilon
}^{\prime }{}^{\ast }\cdot \mathbf{\epsilon }+f_{12}\frac{1}{2}\left(
\mathbf{\hat{k}}_{i}\mathbf{\hat{k}}_{j}+\mathbf{\hat{k}}_{i}^{\prime }%
\mathbf{\hat{k}}_{j}^{\prime }\right) \mathbf{s}^{\prime }{}^{\ast }\cdot
\mathbf{s}-\mathrm{trace}\ ,
\end{eqnarray}%
where $\mathbf{\epsilon (\epsilon ^{\prime })}$ is the polarization for the
initial(final) photon and $\mathbf{\hat{k}(\mathbf{\hat{k}}^{\prime })}$ is
the unit vector in the direction of the initial(final) photon momentum, $%
\mathbf{s}=\mathbf{\hat{k}}\times \mathbf{\epsilon }$ and $\mathbf{s}%
^{\prime }=\mathbf{\hat{k}}^{\prime }\times \mathbf{\epsilon }^{\prime }$.

For unpolarized Compton scattering, the amplitude squared is proportional to
\begin{equation}
\left\vert A\right\vert ^{2}\propto \left\vert S\right\vert ^{2}+\frac{2}{3}%
\left\vert V_{i}\right\vert ^{2}+\frac{4}{3} T_{ij}T_{ij}^*\ ,  \label{Asq}
\end{equation}%
after summing over the initial and final deuteron polarizations. As we shall
see in the following section, the $\left\vert S\right\vert ^{2}$ term starts
to contribute at $\mathcal{O}(Q^0)$ while $\alpha _{0}$ and $\beta _{0}$
contribute at $\mathcal{O}(Q^2)$. The $\left\vert V_{i}\right\vert ^{2}$
term also contributes at $\mathcal{O}(Q^2)$ while the $\left\vert
T_{ij}\right\vert ^{2}$ term only contributes at $\mathcal{O}(Q^3)$. The $%
\left\vert V_{i}\right\vert ^{2}$ term should be included in a $\mathcal{O}%
(Q^2)$ calculation to extract $\alpha _{0}$ and $\beta _{0}$

\section{Unpolarized cross section from EFT}

In the dibaryon formulation of the pion-less effective field theory \cite%
{kaplan,BHvK1,dEFT}, the nucleon field $N$ and the $^{3}S_{1}$-channel
di-baryon field $t_{j}$ are introduced. The effective lagrangian is
\begin{eqnarray}
\mathcal{L} &=&N^{\dagger }\left( iD_{0}+\frac{\mathbf{D}^{2}}{2M_{N}}%
\right) N-t_{j}^{\dagger }\left[ iD_{0}+\frac{\mathbf{D}^{2}}{4M_{N}}-\Delta %
\right] t_{j}  \nonumber \\
&&-y\left[ t_{j}^{\dagger }N^{T}P_{j}N+\mathrm{h.c.}\right]
\end{eqnarray}%
where $P_{i}=\tau _{2}\sigma _{2}\sigma _{i}/\sqrt{8}$ is the $^{3}S_{1}$
two-nucleon projection operators and $y$ is a coupling constant between the
dibaryon and two-nucleon in the same channel. The covariant derivative is $%
\mathbf{D}=\vec{\partial}+ieQ\mathbf{A}$ with $Q=(1+\tau ^{3})/2$ as the
charge operator and $\mathbf{A}$ the photon vector potential. The
nucleon-nucleon scattering amplitude is reproduced by the following choice
of parameters
\begin{equation}
y^{2}=\frac{8\pi }{M_{N}^{2}r^{(^{3}S_{1})}};~~~~\Delta =\frac{2}{%
M_{N}r^{(^{3}S_{1})}}\left( \frac{1}{a^{(^{3}S_{1})}}-\mu \right)
\end{equation}%
where $a^{(^{3}S_{1})}$ is the scattering length, $r^{(^{3}S_{1})}$ is the
effective range, and $\mu $ is the renormalization scale in the power
divergent subtraction scheme \cite{KSW} which conserves gauge symmetry.
Similarly, one can introduce the dibaryon field $S_{i}$\ in the $^{1}S_{0}$
channel.

The magnetic coupling in the lagrangian is
\begin{eqnarray}
\mathcal{L}_{B} &=&\frac{e}{2M_{N}}N^{\dagger }\left( \mu _{0}+\mu _{1}\tau
_{3}\right) \mathbf{\sigma }\cdot \mathbf{B}N+\left[ e\frac{L_{1}}{M_{N}%
\sqrt{r^{(^{1}S_{0})}r^{(^{3}S_{1})}}}t_{j}^{\dagger }S_{3}B_{j}+\mathrm{h.c.%
}\right] \   \nonumber \\
&&-i\frac{e}{M_{N}}\left( \mu _{0}-\frac{L_{2}}{r^{(^{3}S_{1})}}\right)
\varepsilon ^{ijk}t_{i}^{\dagger }B_{j}t_{k},  \label{L2}
\end{eqnarray}%
where $\mu _{0}=(\mu _{p}+\mu _{n})/2$ and $\mu _{1}=(\mu _{p}-\mu _{n})/2$
are the isoscalar and isovector nucleon magnetic moments in units of nuclear
magneton, and $\mathbf{B}$ is the external magnetic field. $L_{1}$ has been
determined by the rate of $np\rightarrow d\gamma $ \cite{CS,dEFT}. The
measured cross section $\sigma ^{\exp }=334.2\pm 0.5$ mb with incident
neutron speed of $2200$ m/s fixes $L_{1}=-4.42$ fm. $L_{2}$ is determined by
the magnetic moment of the deuteron
\begin{equation}
\mu _{d}=2\mu _{0}+\frac{2\gamma L_{2}}{1-\gamma r^{(^{3}S_{1})}}\ ,
\end{equation}%
in units of nuclear magneton, where $\gamma =\sqrt{M_{N}B}=45.703$ MeV with
the deuteron binding energy $B=2.225$ MeV. Fitting to the experimental
value, one finds, $L_{2}=-0.03$ fm.

The relativistic correction of the above nucleon magnetic interaction is the
\textquotedblleft spin-orbit" interaction
\begin{equation}
\mathcal{L}_{SO}=N^{\dagger }i\left[ \left( 2\mu_0-\frac{1}{2}\right)
+\left( 2\mu_1-\frac{1}{2}\right) \tau _{3}\right] \frac{e}{8M_{N}^{2}}%
\mathbf{{\sigma }\cdot \left( \mathbf{D}\times \mathbf{E}-\mathbf{E}\times
\mathbf{D}\right) N\ ,}  \label{SO}
\end{equation}%
where $\mathbf{E}$ is an external electric field. There are also
polarizability contributions%
\begin{eqnarray}
\mathcal{L}_{pol} &=&2\pi N^{\dagger }\left( \alpha _{0}+\alpha _{1}\tau
_{3}\right) \mathbf{E}^{2}N+2\pi N^{\dagger }\left( \beta _{0}+\beta
_{1}\tau _{3}\right) \mathbf{B}^{2}N\   \nonumber \\
&&+\frac{2\pi \alpha _{4}}{M_{N}r^{(^{3}S_{1})}}t_{i}^{\dagger }t_{i}\mathbf{%
E}^{2}+\frac{2\pi \beta _{4}}{M_{N}r^{(^{3}S_{1})}}t_{i}^{\dagger }t_{i}%
\mathbf{B}^{2}  \label{pol}
\end{eqnarray}%
where $\alpha _{0,1}=\left( \alpha _{p}\pm \alpha _{n}\right) /2$ is the
isoscalar (isovector) nucleon electric polarizability, and $\beta
_{0,1}=\left( \beta _{p}\pm \beta _{n}\right) /2$ is the isoscalar
(isovector) magnetic polarizability. $\alpha _{4}$ and $\beta _{4}$ are
unknown two-body currents which will limit the precision of extractions of
the above polarizabilities in deuteron Compton scattering.

In this version of EFT with pions integrated out, the ultraviolet cut-off
scale $\Lambda $ is typically set by the pion mass $m_{\pi }\simeq 140$ MeV,
although the nucleon mass comes in from the nucleon propagator. Following
\cite{dEFT}, we will not distinguish the difference between $M_N$ and $m_\pi$%
. There are some light scales denoted as $Q$ in the system. The inverse
scattering lengths $1/a^{(i)}$ and deuteron internal momentum $\gamma $ are
of order $Q$ while the effective ranges $r^{(i)}$ are counted as $1/Q$. This
counting allows the re-summation of the effective range contributions, which
is a distinctive feature of the dibaryon approach. Note that in another
approach, $r^{(i)}$ is counted as $Q^{0}$ \cite{KSW,chen1}, thus the
effective range contributions are perturbative.

The photon energy $\omega $ scales as $Q^{2}$ at energies comparable and
below the deuteron binding (region I) and as $Q$ at higher energies (region
II) \cite{chen2}. Since the nucleon polarizability contributions are
proportional to $\omega ^{2}$, we will work in region II to accentuate their
effects. Another scale in the problem is the relative momentum $p$ between
the nucleons in the intermediate states. Naively, $p\sim $ $\sqrt{%
M_{N}\omega }$, thus one expects $p\sim \Lambda $ when $\omega \sim 30$ MeV.
However, for the unpolarized deuteron Compton scattering, the pionless
theory still converges well at $\omega \sim 50$ MeV. This is because the
main uncertainty from diagrams with nucleon-nucleon rescattering in
intermediate states are generally suppressed compared to the
non-rescattering ones. Furthermore, the uncertainty is mainly from the $P$
waves and higher, because the S-wave rescattering is described by the
pionless EFT well in this energy range.

Despite the successes of the pionless EFT, it is important that future
experiments can be measured at lower energies such that the theory is
unquestionably under control. We will explore the optimal energy range in
the end of this section.

\begin{figure}[tbp]
\includegraphics[width=4.8in]{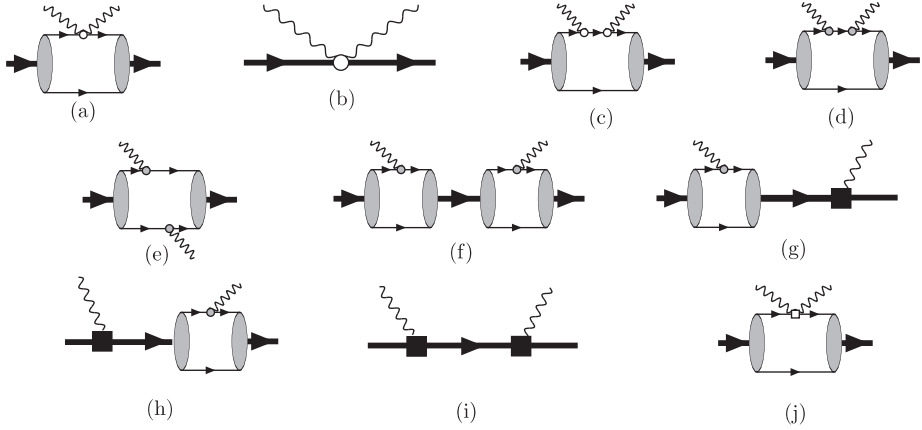}
\caption{{Feynman diagrams for the scalar amplitude of deuteron Compton
scattering. The thick initial and final state arrows represent the deuteron.
The thick arrows in the middle of the diagrams are the dibaryon states in
the $^{3}S_{1}$ and $^{1}S_{0}$ channels. The open circles denote the
electric current interactions, and the shaded circles the magnetic moment
interactions. The solid squares are from electromagnetic counter-terms ($%
L_{1}$ and $L_{2}$). The open square is from the nucleon polarizabilities.}}
\label{fig1}
\end{figure}

Diagrams contributing to the scalar amplitude $S$ (or equivalently $f_{1} $
and $f_{2}$) to $\mathcal{O}(Q^2)$ are shown in Fig.\ref{fig1}. The
gauge-coupling diagrams in 1(a) and 1(b) are of order $Q^{0}$, and of order $%
Q^{1/2}$ in 1(c). The magnetic diagrams in 1(d) is of order $Q$ and of order
$Q^{3/2}$ in 1(e), while those in 1(f)-1(i) are of order $Q^{2}$, with the
solid squares denoting magnetic two-body current $L_{1}$ and $L_{2}$. The
open square in 1(j) is the nucleon polarizability coupling which also
contributes at $Q^{2}$.

The results of the diagrams are expressed in terms of the reduced amplitudes
\begin{equation}
f_{i}=\frac{1}{\left( 1-\gamma r^{(^{3}S_{1})}\right) }\widetilde{f}_{i}\ ,
\end{equation}%
where $i=1-12$. We write
\begin{eqnarray}
\widetilde{f}_{1} &=&g_{1}(\omega )+g_{1}(-\omega )+zh(\omega )\ ,  \nonumber
\\
\widetilde{f}_{2} &=&g_{2}(\omega )+g_{2}(-\omega )-h(\omega )\ ,
\end{eqnarray}%
where%
\begin{eqnarray}
g_{1}(\omega ) &=&\left[ \frac{\gamma \,{{\rho }_{d}}}{2}-\frac{4\,\gamma \,%
}{\sqrt{2-2\,z}\,\omega \,}\tan ^{-1}(\frac{\sqrt{2-2\,z}\,\omega }{%
4\,\gamma })+\frac{2\,\gamma \,\left( \gamma +2\,\sqrt{{\gamma }^{2}-\omega
\,{M_{N}-i\varepsilon }}\right) }{3\,{\left( \gamma +\sqrt{{\gamma }%
^{2}-\omega \,{M_{N}-i\varepsilon }}\right) }^{2}\,}\right. \   \nonumber \\
&&\left. {+}\frac{32\,\pi \,\gamma \,\omega \,\,{M_{N}}\,{{\alpha }_{0}}}{%
e^{2}\,\sqrt{2-2\,z}}\tan ^{-1}(\frac{\sqrt{2-2\,z}\,\omega }{4\,\gamma })%
\right] \ ,
\end{eqnarray}%
and%
\begin{eqnarray}
h(\omega ) &=&\frac{1\,}{15\,{\omega }^{2}\,{{M_{N}}}^{4}\,}\left(
-4\,\gamma \,{\left( {\gamma }^{2}-\omega \,{M_{N}-i\varepsilon }\right) }^{%
\frac{5}{2}}+8\,{\gamma }^{6}+20\,{\gamma }^{4}\,\omega \,{M_{N}}+60\,{%
\gamma }^{2}\,{\omega }^{2}\,{{M_{N}}}^{2}\right.   \nonumber \\
&&\left. +15\,{\omega }^{3}\,{{M_{N}}}^{3}-\gamma \,\sqrt{{\gamma }%
^{2}+\omega \,{M_{N}}}\,\left( 4\,{\gamma }^{4}+28\,{\gamma }^{2}\,\omega \,{%
M_{N}}+39\,{\omega }^{2}\,{{M_{N}}}^{2}\right) \right) \ .
\end{eqnarray}%
and where%
\begin{equation}
g_{2}(\omega )=\frac{32\,\pi \,\gamma \,\omega \,\,{M_{N}}\,{{\beta }_{0}}}{%
e^{2}\,\sqrt{2-2\,z}}\tan ^{-1}(\frac{\sqrt{2-2\,z}\,\omega }{4\,\gamma })+%
\frac{4\,\gamma \,\left( \,\gamma -\sqrt{{\gamma }^{2}-\omega \,{%
M_{N}-i\varepsilon }}\right) \,{{{\mu }_{1}}}^{2}}{3\,{{M_{N}}}^{2}}%
+k(\omega )\ ,
\end{equation}%
and%
\begin{equation}
k(\omega )=-\frac{\gamma \,\left[ {\omega \,{L_{1}}\,{M_{N}}+2\,\left(
\gamma -\sqrt{{\gamma }^{2}-\omega \,{M_{N}-i\varepsilon }}\right) \,{{\mu }%
_{1}}}\right] ^{2}}{3\,{{M_{N}}}^{2}\,\left( \frac{1}{a^{(^{1}S_{0})}}-\sqrt{%
{\gamma }^{2}-\omega \,{M_{N}-i\varepsilon }}+\frac{r^{(^{1}S_{0})}\,}{2}%
\left( {\gamma }^{2}-\omega \,{M_{N}}\right) \right) }\ .  \label{k}
\end{equation}

These results agree with those obtained in the previous calculations \cite%
{dEFT,rupak}. If the effective range is counted as order $Q^{0}$ instead of $%
Q^{-1}$ then the results of Ref. \cite{rupak} are reproduced. Also, the
results of Ref. \cite{dEFT} are reproduced up to recoil effects which can be
safely neglected.


\begin{figure}[tbp]
\includegraphics[width=5.5in]{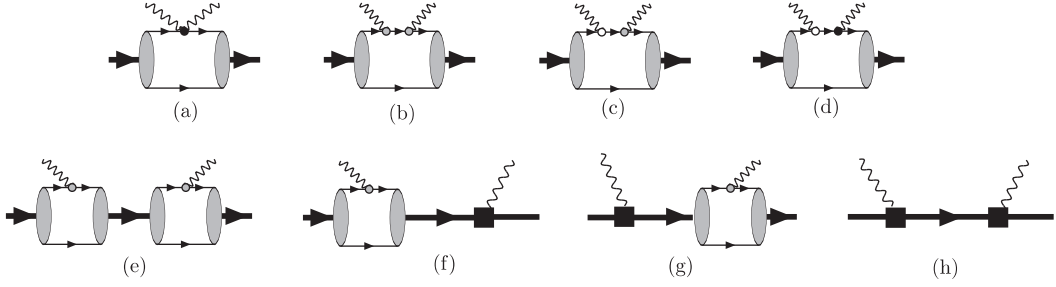}
\caption{{Feynman diagrams for the vector amplitude of deuteron Compton
scattering. The solid circles denote the spin-orbit intercation in Eq. (%
\protect\ref{SO}). Other symbols are the same as described in Fig. 1.}}
\label{fig2}
\end{figure}

The new contributions we calculate are shown in Fig. \ref{fig2}. Those are
diagrams contributing to the vector amplitude $V$ or, equivalently, the
amplitudes $f_{3,...,6}$. The solid circle denotes the spin-orbit coupling
in Eq. (\ref{SO}). The leading diagrams are 2(a)-2(c) which are of order $Q$%
. It is interesting to note that the relativistic correction (the spin-orbit
interaction in diagram 2(a)) contributes at leading order. Diagram 2(d) is
of order $Q^{3/2}$, while diagrams 2(e)-2(h) are of order $Q^{2}$. Naively,
if we want to calculate the cross section to order $Q^{2}$, these orders $%
Q^{3/2}$ and $Q^{2}$ diagrams are not needed according to Eq. (\ref{Asq}).
We still choose to include these contributions because there is an
enhancement factor ${\mu }_{1}^{2}\sim 5.5\sim {1/Q}$ in them. The diagrams
give%
\begin{eqnarray}
\widetilde{f}_{3} &=&g_{3}(\omega )-g_{3}(-\omega )+2q(\omega )\ ,  \nonumber
\\
\widetilde{f}_{4} &=&g_{4}(\omega )-g_{4}(-\omega )\ ,  \nonumber \\
\widetilde{f}_{5} &=&0\ ,  \nonumber \\
\widetilde{f}_{6} &=&-q(\omega )\ ,
\end{eqnarray}%
where%
\begin{equation}
g_{3}=\frac{2\,\gamma \left( 2\,{{\mu }_{0}}+2\,{{\mu }_{1}-1}\right) }{{%
M_{N}}}\left[ \frac{\,1}{\sqrt{2-2\,z}\,}\tan ^{-1}(\frac{\sqrt{2-2\,z}%
\,\omega }{4\,\gamma })\right] \ .
\end{equation}%
\begin{eqnarray}
q(\omega ) &=&-\frac{\left( {{\mu }_{0}}+{{\mu }_{1}}\right) }{3\,\omega \,{{%
M_{N}}}^{3}\,\sqrt{{\gamma }^{2}+\omega \,{M_{N}}}}\left[ -2\,{\gamma }%
^{5}+2\,{\gamma }^{3}\,\sqrt{{\gamma }^{2}+\omega \,{M_{N}-i\varepsilon }}%
\,\left( 2\,\gamma -\sqrt{{\gamma }^{2}-\omega \,{M_{N}-i\varepsilon }}%
\right) \right.   \nonumber \\
&&+\gamma \,\omega \,{M_{N}}\,\left( -7\,{\gamma }^{2}+3\,\gamma \,\sqrt{{%
\gamma }^{2}+\omega \,{M_{N}}}+2\,\sqrt{{\gamma }^{4}-\omega ^{2}\,{%
M_{N}^{2}-i\varepsilon }}\,\right)   \nonumber \\
&&\left. +{\omega }^{2}\,{{M_{N}}}^{2}\,\left( -5\,\gamma +3\,\sqrt{{\gamma }%
^{2}+\omega \,{M_{N}}}\right) \,\right] \ ,
\end{eqnarray}%
and%
\begin{equation}
g_{4}(\omega )=\frac{3}{2}k(\omega )-\left( {{\mu }_{0}^{2}}+{{\mu }_{1}^{2}}%
\right) \left[ \frac{\omega }{{M_{N}}}+\frac{2\gamma }{{M_{N}^{2}}}\sqrt{{%
\gamma }^{2}-\omega \,{M_{N}-i\varepsilon }}\right] \ ,
\end{equation}%
with $k(\omega )$ as given in Eq. (\ref{k}).

As noted above, the tensor amplitude $T$ only contributes to the cross
section at ($Q^{3}$). At the same order, unknown two-body currents $\alpha
_{4}$ and $\beta _{4}$ defined in Eq. (\ref{pol}) also contribute. Thus we
just calculate the unpolarized Compton scattering to order $Q^{2}$. This
will allow us to extract the nucleon electric polarizability $\alpha _{0}$
with $\sim 30\%$ theoretical uncertainty.

Combining the above results, the differential cross section for unpolarized
Compton scattering is
\begin{eqnarray}
\left. \frac{d\sigma }{d\Omega }\right\vert _{cm} &=&\frac{\alpha ^{2}}{%
6(\omega +2M_{N})^{2}}\text{Re}\left[ 3(1+z^{2})(f_{1}^{\ast
}f_{1}+f_{2}^{\ast }f_{2})+12zf_{1}^{\ast }f_{2}\right.  \nonumber \\
&&+2(3-z^{2})(f_{3}^{\ast }f_{3}+f_{4}^{\ast }f_{4})+8zf_{3}^{\ast
}f_{4}+4(1+3z^{2})(f_{5}^{\ast }f_{5}+f_{6}^{\ast }f_{6})  \nonumber \\
&&+8z(3+z^{2})f_{5}^{\ast }f_{6}+8(1+z^{2})(f_{3}^{\ast }f_{6}+f_{4}^{\ast
}f_{5})+16z(f_{3}^{\ast }f_{5}+f_{4}^{\ast }f_{6})]\ ,
\end{eqnarray}%
where $\alpha =1/137$ is the fine structure coupling constant and $z=\cos
\theta =\mathbf{\hat{k}}\cdot \mathbf{\hat{k}}^{\prime }.$


\begin{figure}[tbp]
\includegraphics[width=3.3in]{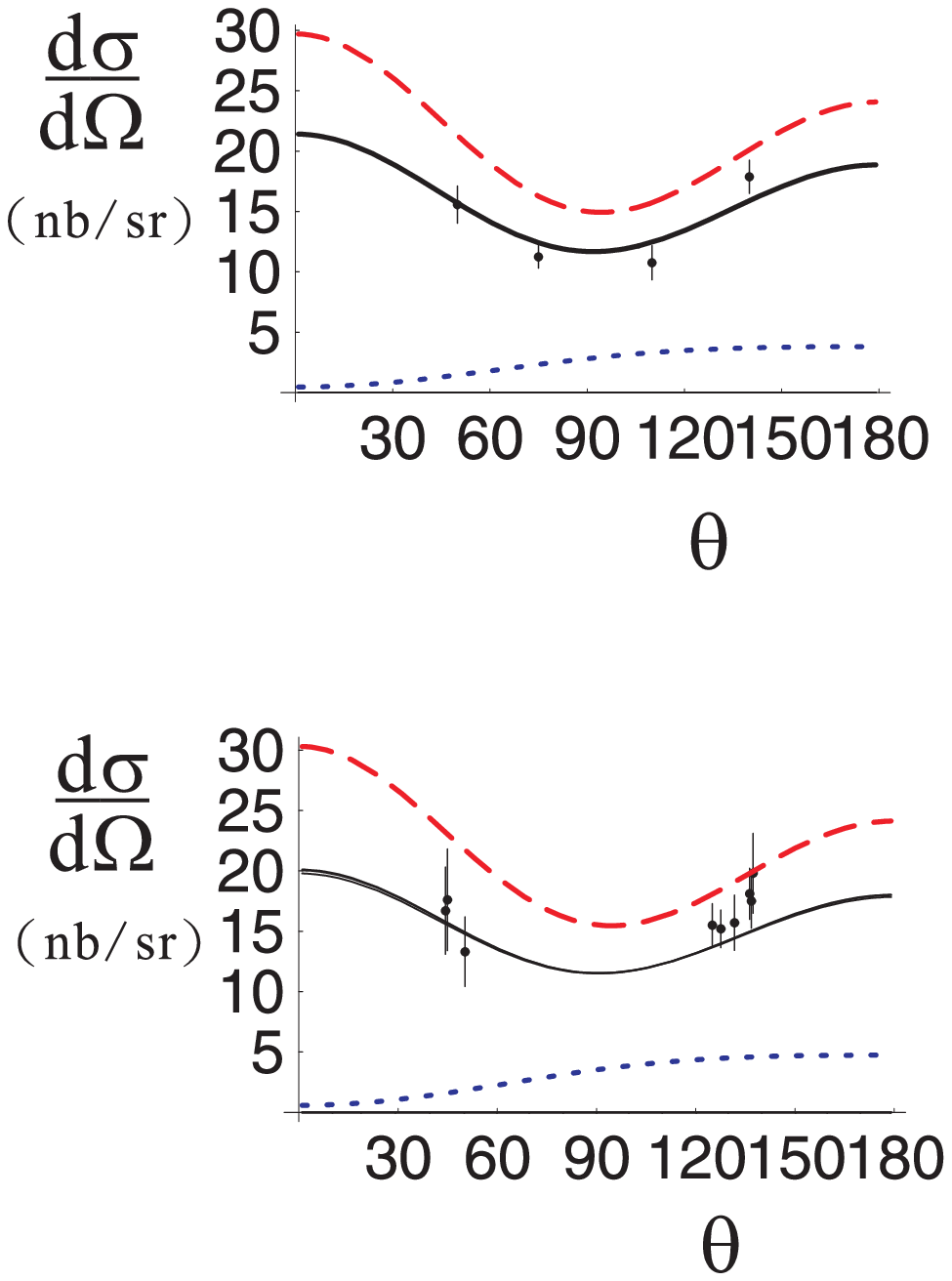}
\caption{{Differential cross section for $\protect\gamma d\rightarrow
\protect\gamma d$ with incident photon energy $\protect\omega =49$ MeV
(upper plot) and $\protect\omega =$55 MeV (lower plot). The dashed curves
are $\mathcal{O}(Q^2)$ results with no nucleon polarizabilities, $\protect%
\alpha _{0}=\protect\beta _{0}=0$. The solid curves are results using
polarizabilities $\protect\alpha _{0}=10\protect\beta _{0}=1.2\times 10^{-3}$
fm$^{3}$, calculated in leading order chiral perturbation theory
\protect\cite{meissner}. The dotted curves are the vector amplitudes
contributions alone. The 49 MeV data is from Ref. \protect\cite{Illinois},
while the 55 MeV data is from Ref. \protect\cite{Lund}. }}
\label{fig3}
\end{figure}

In Fig. \ref{fig3}, we show the differential cross section for $\omega =49$
MeV (upper plot) and $\omega =$55 MeV (lower plot). The $49$ MeV data is
measured at Illinois while the 55 MeV data is actually a superposition of
the 54.6, 54.9 and 55.9 MeV data from the Lund experiment. In both plots,
the dashed curves are $\mathcal{O}(Q^{2})$ results with no nucleon
polarizabilities, $\alpha _{0}=\beta _{0}=0$. The solid curves are results
using nucleon polarizabilities calculated in leading order chiral
perturbation theory (ChPT) \cite{meissner},
\begin{equation}
\alpha _{0}=10\beta _{0}=\frac{5g_{A}^{2}e^{2}}{192\pi ^{2}f_{\pi
}^{2}m_{\pi }}=1.2\times 10^{-3}\text{ fm}^{3}\ .  \label{ChPT}
\end{equation}%
The dotted curves are the contributions of the vector amplitudes which were
not included in the previous calculations \cite{chen2,dEFT,rupak}. These
effects are $\sim $20\% and are as large as the nucleon polarizability
contributions in the backward angles.


\begin{figure}[tbp]
\includegraphics[width=6.3in]{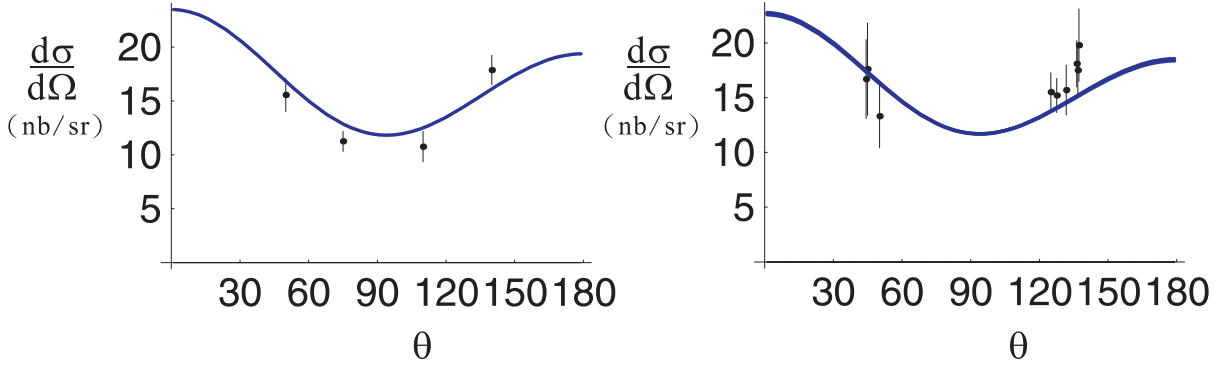}
\caption{{Differential cross section for $\protect\gamma d\rightarrow
\protect\gamma d$ with incident photon energy $\protect\omega =49$ MeV (left
plot) and $\protect\omega =$55 MeV (right plot). The leading-order ChPT
values $\protect\alpha _{0}=10\protect\beta _{0}=1.2\times 10^{-3} $ fm$^{3}$
are used and the effective range parameter is treated as order $Q^{0} $. }}
\label{fig4}
\end{figure}

In Fig. \ref{fig4} we plot the cross section using the counting employed in
Ref. \cite{rupak} and the value of Eq. (\ref{ChPT}). We set $1/a^{(i)}\sim
\omega \sim Q$, $r^{(i)}\sim Q^{0}$ (such that the effective range
contributions are treated perturbatively) and expand our amplitudes in $Q$.
The leading order ChPT value of nucleon polarizabilities also gives a good
description of data. Setting the vector amplitude to be zero, our result
coincides with that of Fig. 2 of Ref. \cite{rupak}.


\begin{figure}[tbp]
\includegraphics[width=6.3in]{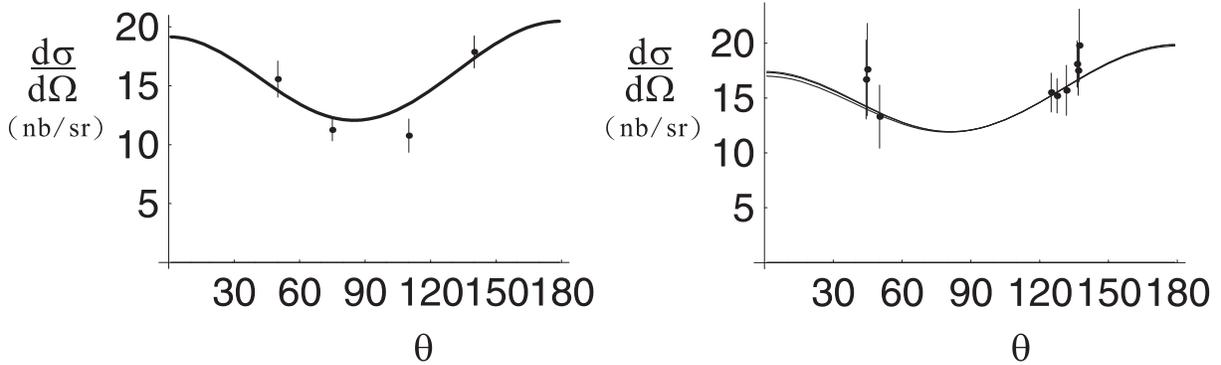}
\caption{{Differential cross section for $\protect\gamma d\rightarrow
\protect\gamma d$ with incident photon energy $\protect\omega =49$ MeV (left
plot) and $\protect\omega =$55 MeV (right plot). The best fit value $\protect%
\alpha _{0}=12.3$ and $\protect\beta _{0}=5.0$ (in units of fm$^{-4}$) are
used and all data points are assumed to be independent in the fitting. }}
\label{fig5}
\end{figure}

Assuming the data measured at $\omega =$49, 54.6, 54.9 and 55.9 MeV are
independent, a $\chi ^{2}$ fit gives $\alpha _{0}=12.3\pm 1.4$ and $\beta
_{0}=5.0\pm 1.6$ (note that theoretical uncertainty is not included) in the
dibaryon formalism. The best fit curves are plotted in Fig. \ref{fig5}. In
the counting that treats the effective range perturbatively, the $\chi ^{2}$
fit gives $\alpha _{0}=14.2\pm 2.1$ and $\beta _{0}=9.3\pm 2.5$ (theoretical
uncertainty is not included) with very similar curves as those shown in Fig. %
\ref{fig5}. The difference between these two counting schemes is of higher
order, thus it appears that while the fit values of $\alpha _{0}$ are quite
close to the one predicted by leading order ChP{}T, the error on $\beta _{0}$
is quite large using the current data. These results can be compared with $%
\alpha _{0}=8.4\pm 3.4$ and $\beta _{0}=8.9\pm 4.3$ from Ref. \cite{rupak},
where the contribution of the vector amplitude has been neglected. Thus the
vector amplitude has significant effect on a reliable extraction of $\alpha
_{0}$.


\begin{figure}[tbp]
\includegraphics[width=3.3in]{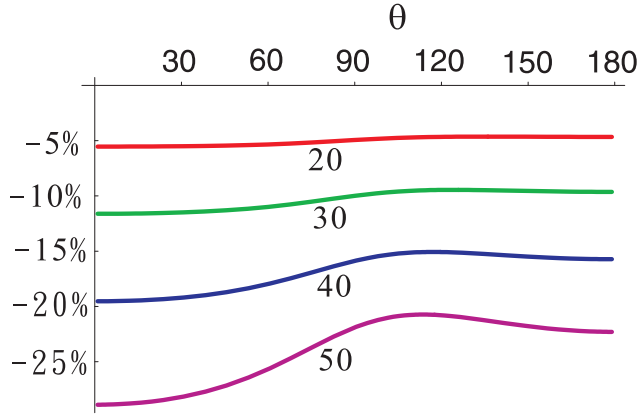}
\caption{{Nucleon polarizability effects on Compton scatterong cross section
with $\protect\alpha _{0}=10\protect\beta _{0}=12$ (in units of 10$^{-4}$ fm$%
^{3}$). The curves, from top to bottom, are for photon energy $\protect%
\omega =20$, 30, 40 and 50 MeV, respectively. }}
\label{fig6}
\end{figure}

Future high energy, high precision experiments can take advantage of the
theoretically clean pionless EFT by performing the measurement at lower
energies. However, the nucleon polarizability contributions become less
important at lower energies. To study the best energy range for future
experiments, we show the nucleon polarizability effects in the Compton
scattering cross section for $\alpha _{0}=10\beta _{0}=12$ (the leading
order ChPT value) in Fig. \ref{fig6}. The curves, from top to bottom, are
for photon energy $\omega =20$, 30, 40 and 50 MeV, respectively. Given the
unknown $\mathcal{O}(Q^{5/2})$ effect is about 5\% for a series with
expansion parameter $\sim 1/3$ and the theory converges better at $\lesssim $%
30 MeV, we recommend the best energy range to perform the deuteron Compton
measurement is 25-35 MeV. From this figure, we see that a measurement at 30
MeV with 3\% error will constrain $\alpha _{0}$ with $3\times 10^{-4}$ fm$%
^{3}$ experimental error, which is comparable to the expected theoretical
error.

\section{Conclusion}

We have computed the unpolarized Compton scattering in dibaryon effective
field theory at $\mathcal{O}(Q^2)$. The vector amplitude, which contributes
to the cross section at $\mathcal{O}(Q^2)$ but was not included in previous
pionless EFT calculations, was found to affect the extraction of nucleon
electric polarizability by more than $50\%$ at 49 MeV. We recommend future
high precision deuteron Compton scattering experiments be measured at 25-35
MeV photon energy for appreciable nucleon polarizability effects and
controllable theoretical higher-order effects. More specifically, a
measurement at 30 MeV with 3\% error will constrain $\alpha _{0}$ with $%
3\times 10^{-4}$ fm$^{3}$ experimental error, which is comparable to the
expected theoretical error.

This work was supported by the U. S. Department of Energy via grant
DE-FG02-93ER-40762 and by the National Science Council of Taiwan, ROC. JWC
thanks Paulo Bedaque for organizing the Summer of Lattice Workshop 2004 at
Lawrence Berkeley Laboratory where part of this research was completed.

\end{document}